\documentclass[preprint,aps,floatfix]{revtex4}
\usepackage{graphicx,epsfig,amssymb,amsmath,array}
\usepackage{relsize,placeins,float}
\usepackage[usenames]{color}

\begin{document}
\title{Fractional damping induces resonant behavior in the Duffing oscillator}
\author{Mattia Coccolo}
\affiliation{Nonlinear Dynamics, Chaos and Complex Systems Group, Departamento de F\'{i}sica,
Universidad Rey Juan Carlos, Tulip\'{a}n s/n, 28933 M\'{o}stoles (Madrid), Spain}
\author{Jes\'{u}s M. Seoane}
\affiliation{Nonlinear Dynamics, Chaos and Complex Systems Group, Departamento de F\'{i}sica,
Universidad Rey Juan Carlos, Tulip\'{a}n s/n, 28933 M\'{o}stoles (Madrid), Spain}
\author{Miguel A.F. Sanju\'{a}n}
\affiliation{Nonlinear Dynamics, Chaos and Complex Systems Group, Departamento de F\'{i}sica,
Universidad Rey Juan Carlos, Tulip\'{a}n s/n, 28933 M\'{o}stoles (Madrid), Spain}
\date{\today}

\begin{abstract}

The interaction between the fractional order parameter and the damping parameter can play a relevant role for introducing different dynamical behaviors in a physical system. Here, we study the Duffing oscillator with a fractional damping term. Our findings show that for certain values of the fractional order parameter, the damping parameter, and the forcing amplitude high oscillations amplitude can be induced. This phenomenon is due to the appearance of a  resonance in the Duffing oscillator only when the damping term is fractional. 

\end{abstract}

\maketitle

\section{Introduction}\label{sec:introduction}

Nonlinear oscillators are dynamical systems that exhibit oscillatory behaviors but do not follow a linear relationship between the restoring force and the displacement. These systems often arise in various scientific and engineering fields, and their study has profound implications in understanding complex phenomena and designing efficient devices. When considering damping in nonlinear oscillators, a commonly used approach is to introduce a fractional derivative in the damping term. The use of fractional derivatives in damping terms provides a powerful tool to capture the memory effects and non-local behavior observed in certain physical systems. Unlike the traditional integer-order derivatives that describe instantaneous rates of change, fractional derivatives involve a non-integer order and account for past history in the damping process. This allows us a more accurate representation of damping in systems with memory and hysteresis as occurs in some properties of complex materials. The presence of a fractional derivative damping term in nonlinear oscillators can lead to novel behaviors such as sub-diffusive or super-diffusive damping, where the decay of oscillations is slower or faster than in traditional damping scenarios. Moreover, the inclusion of the fractional derivatives can alter the stability regions, bifurcation boundaries, and resonance responses of the oscillatory system, introducing rich and intricate dynamics. The study of nonlinear oscillators with fractional derivative in the damping term  finds applications in disciplines such as mechanical engineering, physics, robotics, control systems, and even social sciences \cite{Boroviec,Dafermos,Lu,Chellaboina,De,Elliott,Horr,Ding,Li,Strogatz}. Understanding the behavior and control of these oscillators is crucial for optimizing energy dissipation, enhancing stability, and designing systems with specific response characteristics.

On the other hand, nonlinear resonances~\cite{Sanjuan} is a fascinating phenomenon that occurs when a system, subjected to an external force or perturbation, exhibits a response that is not proportional to the input. Unlike linear resonance, where the response is directly proportional to the forcing frequency, nonlinear resonances arise due to the intricate interplay between multiple frequencies and energy transfer mechanisms within the system, as in our case the fractional order parameter, the damping parameter and the forcing amplitude. In nonlinear resonances, the system's response can display a wide range of complex behaviors, including the generation of harmonics, sub-harmonics, chaotic oscillations, and bifurcations. As stated before, the fractional derivatives are the key to explore and modeling materials with memory effects and fractal geometry or diffusion processes, among others. In fact, fractional derivatives can play a crucial role on them. Therefore, it is relevant to analyze when the fractional order parameter can trigger complex behaviors and nonlinear resonance phenomena.

Here, we analyze the Duffing oscillator with a fractional derivative in its damping term as in Ref.~\cite{Coccolo_fr2}. Specifically, we focus our study in the numerical analysis of the fractional order parameter effects on the oscillations amplitude of the system.  We have found that the fractional order parameter can induce resonance phenomena. Firstly, we analyze, for smaller forcing amplitude values, beyond a certain threshold, the appearance of resonance peaks that are triggered by the high damping parameter values and specific values of the fractional order parameter. These high amplitude oscillations are related with a complex dynamics of the oscillator in the phase space. Then, we find a region of resonances for higher forcing amplitude values. Consequently, we characterize and analyze the appearance of  resonance phenomena induced by the effect of the three main ingredients: high forcing amplitude and damping parameter values, and some specific values of the fractional order parameter. It is relevant to emphasize that these complex behaviors and resonance phenomena are an effect of the fractional order parameter, and they only appear when a fractional derivative damping term exists. 

The organization of this paper is as follows.  In Sec.~\ref{s:model}, we introduce the fractional Duffing oscillator and its dynamics. In Sec.~\ref{section_3}, we numerically present the effects on the oscillations amplitude of the fractional order parameter in which a resonant-like behavior is uncovered. The impact of the forcing amplitudes due to the fractional order parameter is analyzed in Sec.~\ref{section_4}, where we numerically characterize the birth of this resonance-like behavior. Conclusions and a discussion of the main results of this paper are presented in Sec.~\ref{sec:conclusions}.

\section{Model Description}\label{s:model}

\begin{figure}[htbp]
  \centering
   \includegraphics[width=16.0cm,clip=true]{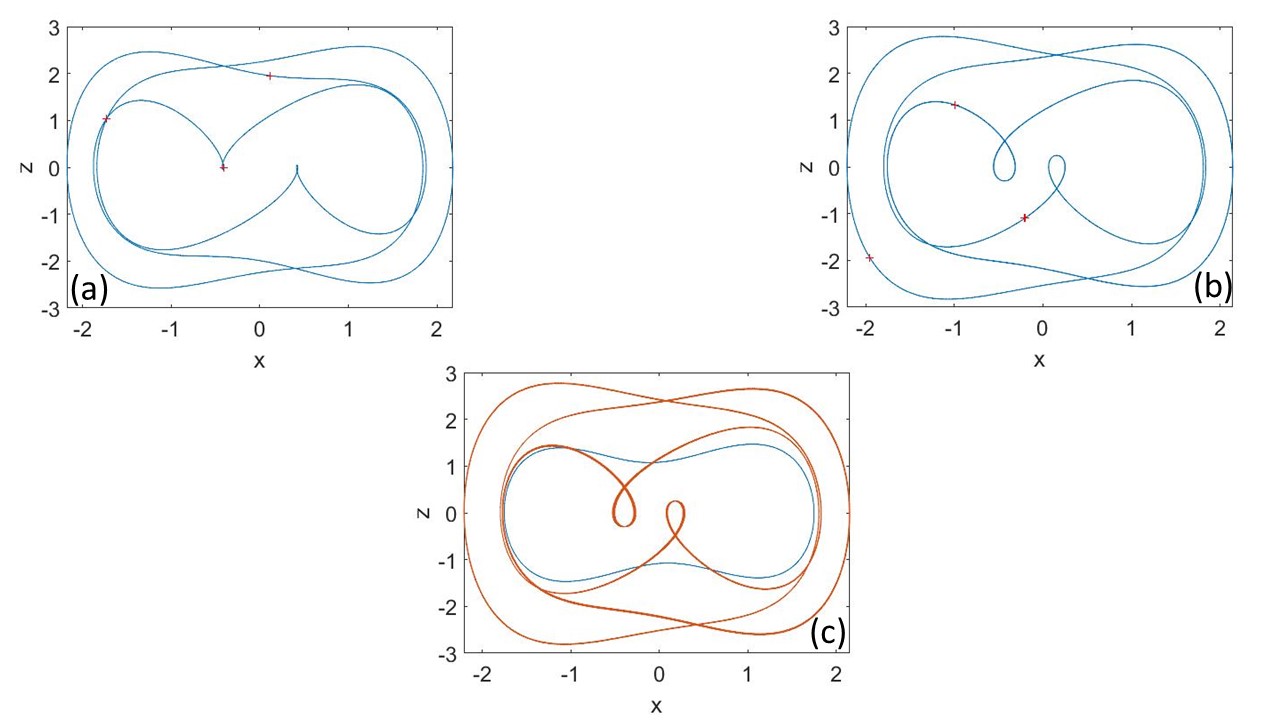}
   \caption{In this figure, we have plotted the asymptotic behaviors of the system trajectory in phase space for (a) $\mu=0.6$ and $\alpha=0.3$, (b) $\mu=0.9$ and $\alpha=0.1$ and (c) $\mu=0.9$ and $\alpha=0.1$ (in red) and $\alpha=0.3$ (in blue). We fix the forcing amplitude $F=1.5$ for all cases. Note the key role of the $\alpha$ and $\mu$ parameters in the amplitude of the oscillations.}
\label{fig:0}
\end{figure}

 We study a Duffing oscillator with a fractional derivative of fractional order parameter $\alpha$ instead of the first derivative, i.e., $\dot{x}(t) \to D^{\alpha}x(t)=d^{\alpha}x(t)/dt^{\alpha}$ as a damping term.  Then, the Duffing oscillator with a fractional damping term is given by
 \begin{equation}
     \frac{d^2x}{dt^2}+\mu\frac{d^{\alpha}x}{dt^{\alpha}}-x+x^3=F\cos{\omega t},
 \end{equation}
 where $\mu$ is the damping parameter, $F$ and $\omega$ are the forcing amplitude and forcing frequency, respectively, and $\alpha$ is the fractional order parameter. In order to integrate the system, we use the {\it Gr\"{u}nwald-Letnikov} fractional derivative, as in previous works \cite{Coccolo_fr,Coccolo_fr2}. Consequently, the system is composed by a set of three fractional differential equations that yield

\begin{equation}\label{model}
\left\lbrace\begin{array}{l}
 D^{\alpha}x=y, \medskip\\
 D^{1-\alpha}y=z, \medskip\\
 Dz= \mu y + F\cos{(\omega t)}+x-x^3,
\end{array}\right.
\end{equation}
where $z$ is an auxiliary component coming from the transformation into a fractional order system. The integration step used is $\pi/2000$, and the maximum time value is $t=300$ that is large enough for the system to reach the steady state.  We have carried out numerical simulations with several initial conditions and, although the results are numerically different, the general trends of the curves are qualitatively similar. So, we carry on the study with the initial condition $(x_{0}, y_{0}, z_{0}) = (0.2, 0.3, 0.0)$, as in Ref.~\cite{Syta, Coccolo_fr2}. We expect that our conclusions are of general validity and not specific for the considered boundary conditions. In Fig.~\ref{fig:0}, we show some examples of trajectories of the system and the impact of the $\alpha$ value on its dynamics, when the $\mu$ parameter varies. In fact, according to Fig.~\ref{fig:0}(a), where $\mu=0.6$, and Fig.~\ref{fig:0}(b), where $\mu=0.9$, we have to change the $\alpha$ parameter from $\alpha=0.3$ to $\alpha=0.1$, respectively, if we want to observe similar oscillations amplitudes. In Fig.~\ref{fig:0}(c), we compare trajectories for $\mu=0.9$ and $\alpha=0.3$ (in blue) and for $\mu=0.9$ and $\alpha=0.1$ (in red). Having traced more trajectories than those shown, we have seen that the phenomenon is complex and worthy of study.

\section{Fractional order parameter effects on the oscillations amplitude}\label{section_3}

\begin{figure}[htbp]
  \centering
   \includegraphics[width=16.0cm,clip=true]{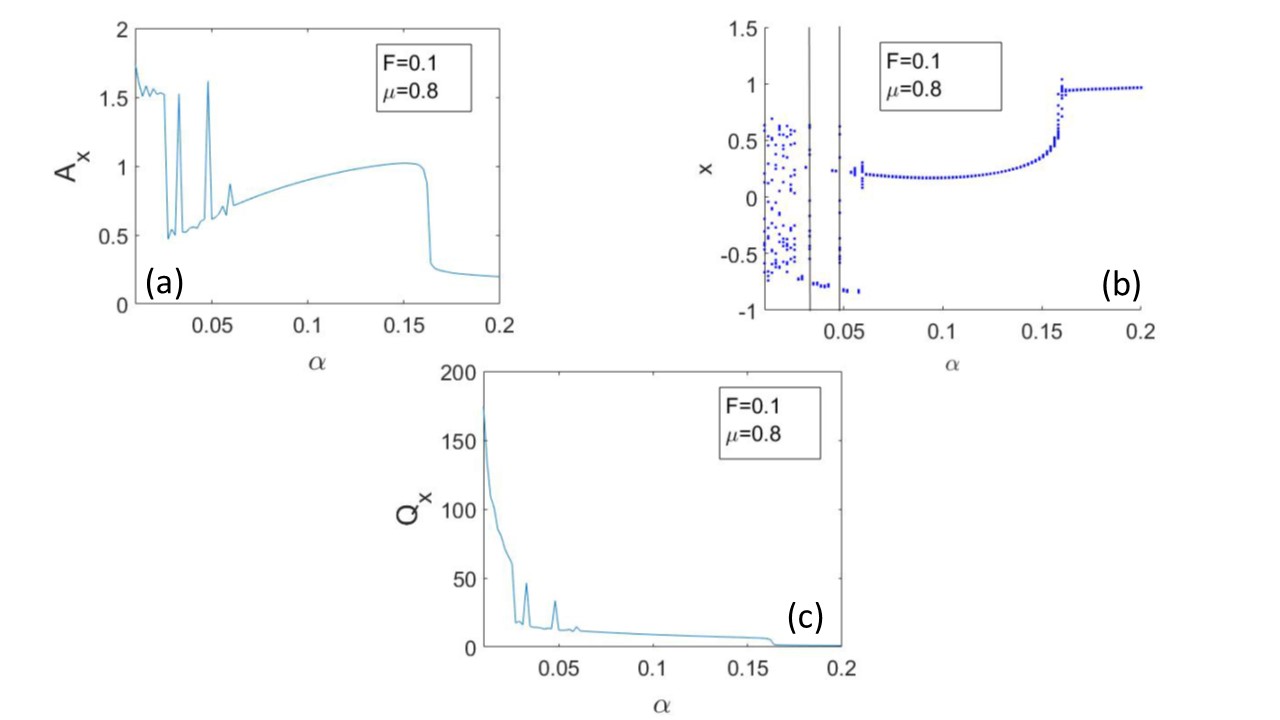}
   \caption{We plot here (a) the oscillations amplitude, (b) the asymptotic behaviors diagram and (c) the $Q-$factor for $F=0.1,\mu=0.8$.  The two resonance peaks can be recognise in the panels.}
\label{fig:1}
\end{figure}

As we previously mentioned, some preliminary computations about the resonance induced by the fractional derivative in the Duffing damping term have been done in~\cite{Coccolo_fr2}. The authors have disclosed that for a high value of the damping parameter $\mu=0.8$ and for distinctive values of the fractional order parameter $\alpha$, some interesting peaks pop up when the forcing amplitude is large enough, as shown in Fig.~\ref{fig:1}. Also, for higher $\alpha$ values we have seen that the oscillations amplitude do not show those kind of peaks and that the curve stays smooth under an oscillations amplitude the of value $0.5$. We can see that two particular peaks stand out in the amplitude and the $Q-$factor plots, Fig.~\ref{fig:1}(a) and Fig.~\ref{fig:1}(c), respectively. However, this fact was just mentioned, though not analyzed in detail. As stated before, the $Q-$factor provides an idea of how much the signal is amplified by a certain parameter, in our case $\alpha$. We calculate it by computing the sine and cosine components:
\begin{equation}\label{eq:3}
    B_s=\frac{2}{nT}\int^{nT}_0{x(t)\sin{\omega t}dt}\quad\text{and}\quad B_c=\frac{2}{nT}\int^{nT}_0{x(t)\cos{\omega t}dt},
\end{equation}
where $T=2\pi/\omega$ and $n$ is an integer. Then, we can find the dependence on $\alpha$ with
\begin{equation}\label{eq:4}
    Q=\frac{\sqrt{B^2_s+B^2_c}}{\alpha}.
\end{equation}

\begin{figure}[htbp]
  \centering
   \includegraphics[width=14.0cm,clip=true]{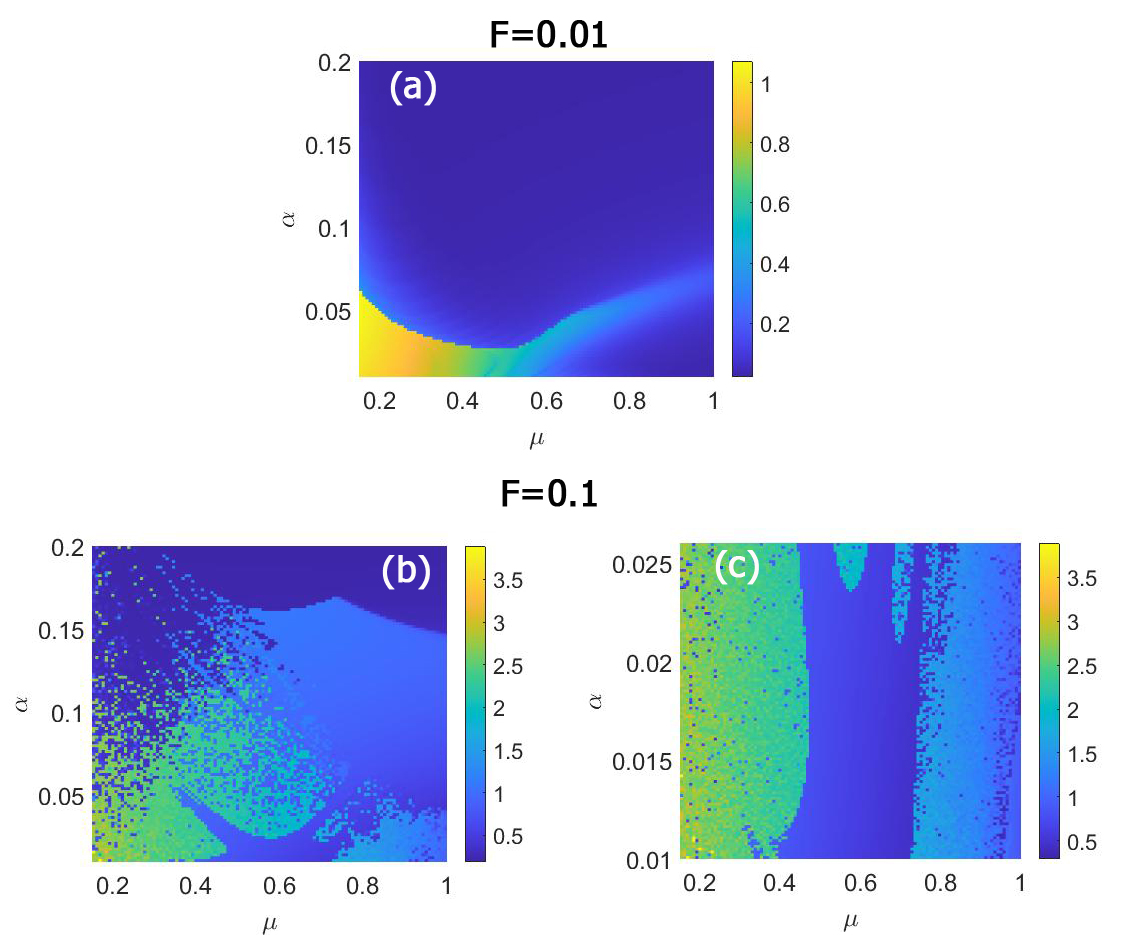}
   \caption{The figure shows the oscillations amplitude for $F=0.01$ (a) and $F=0.1$ (b). Panel (c) is a zoom of panel (b). Here it is possible to appreciate how the difference in the amplitude forcing complicates the topology of the parameter set. }
\label{fig:2}
\end{figure}

Those peaks are related with fractional order parameter values that suddenly give birth to aperiodic orbits, as we can see in Fig.~\ref{fig:1}(b). They are resonance-like peaks for the $\alpha$ parameter, as confirmed by the $Q-$factor, but it is worth of study the conditions for them to appear and the parameter values for which the gain in amplitude is higher. Our objective now is to analyze in depth this resonant behavior. For this purpose,  we plot the gradient of the amplitude in the parameter set $\alpha-\mu$ for two values of the forcing amplitude $F=0.01$ and $F=0.1$ in Fig.~\ref{fig:2}. In the first case, Fig.~\ref{fig:2}(a), we can see that the gradient in the parameter set is smooth, the oscillations amplitude is not that high and the higher oscillations amplitude zone is very easy to spot. The only interesting thing is the intermediate oscillations amplitude smudge for $\mu\gtrsim 0.6$. Even more interesting are the other two panels  where Figs.~\ref{fig:2}(c) is a zoom of Fig.~\ref{fig:2}(b). When we reach the forcing amplitude value $F=0.1$, the parameter set becomes more complicated and the high and low oscillations amplitude intermingle.

\begin{figure}[htbp]
  \centering
   \includegraphics[width=16.0cm,clip=true]{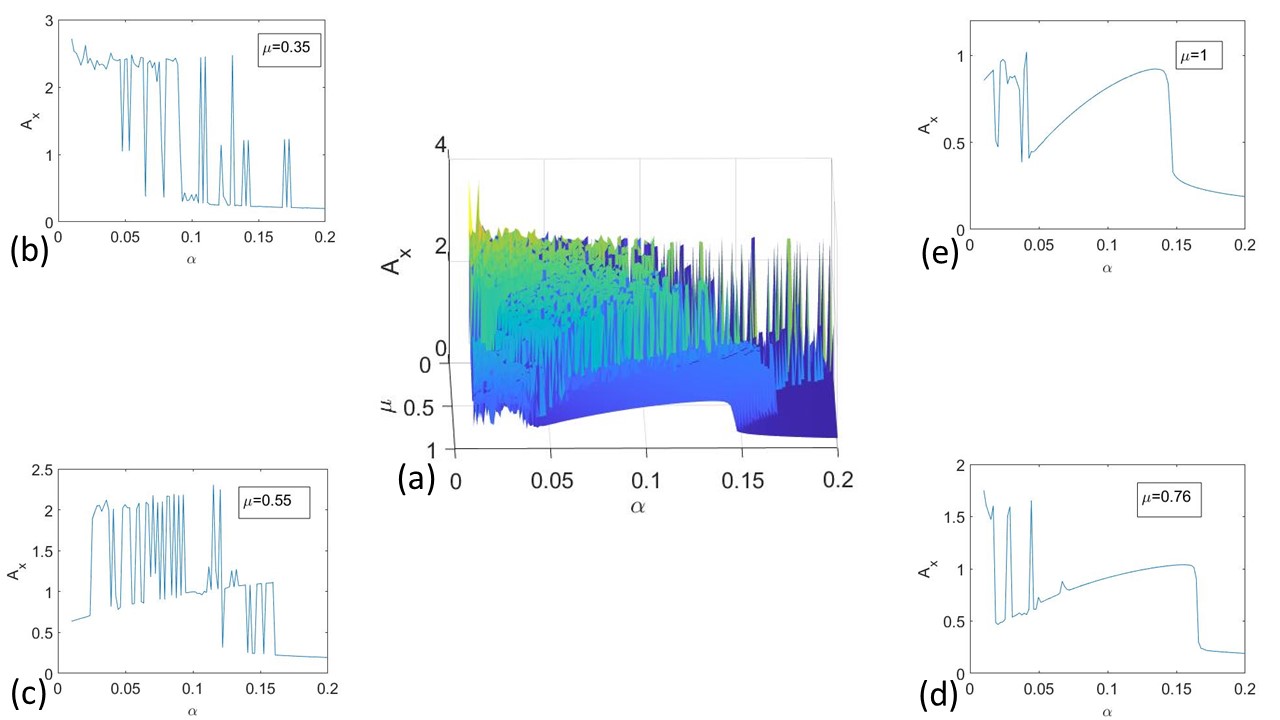}
   \caption{In panel (a), we show the $3D$ version of Fig.~\ref{fig:2}(b). The other panels display $4$ slices of the panel (a) along the $\alpha$ axis and for (b) $\mu=0.35$, (c) $\mu=0.55$, (d) $\mu=0.76$, and (e) $\mu=1$. Here,  the formation of the central structure is clear. In fact, in panel (c) it is possible to see how the high oscillations amplitude of panel (b) decreases and start to form the central structure of panels (d) and (e).}
\label{fig:3}
\end{figure}

 From values $\mu\gtrsim 0.4$, the structure that can be spotted in Fig.~\ref{fig:1}(a) and that ends at $\alpha\approx0.15$, starts to take form with lots of the resonance peaks, in which the oscillations span both potential wells. In fact, in Fig.~\ref{fig:3} we show the oscillations amplitude plots for different $\mu$ values as a section of the $3-D$ figure. There, we can see that for $\mu=0.35$ there is no sign of the structure, while it is visible for $\mu=0.55$. Clearly, for higher $\mu$ values it becomes even more visible. In the $3-D$ figure, it is possible to discern the built-up of such structure and how the higher oscillations amplitude peaks diminish when $\mu$ grows. This structure is the result of the higher oscillations amplitude  being attenuated by the dissipation term, so that the peaks are the consequence of the fractional order parameter feat interacting with the forcing amplitude. Hence, it seems that they are fractional-induced amplitude peaks that behave like a resonance when only the fractional order parameter varies. In fact, if we compute the $Q-$factor for the parameter $\mu$ the peaks do not show up in the plot.

In Fig.~\ref{fig:3}(e), we can see that the trajectories become bounded to one well when the dissipation is too strong, $\mu=1$, for the fractional order parameter to maintain the interwell oscillations.  Figure~\ref{fig:2}(c) has been plotted to have a better understanding of the higher amplitude oscillations zone. In particular of the cross like patch centered at $\mu=0.6$. Here, we can find a region of small oscillations amplitude embedded inside a high oscillations amplitude region. It is precisely there, where the already mentioned structure, well recognisable in Fig.~\ref{fig:1}(a), starts to take form.

\begin{figure}[htbp]
  \centering
   \includegraphics[width=16.0cm,clip=true]{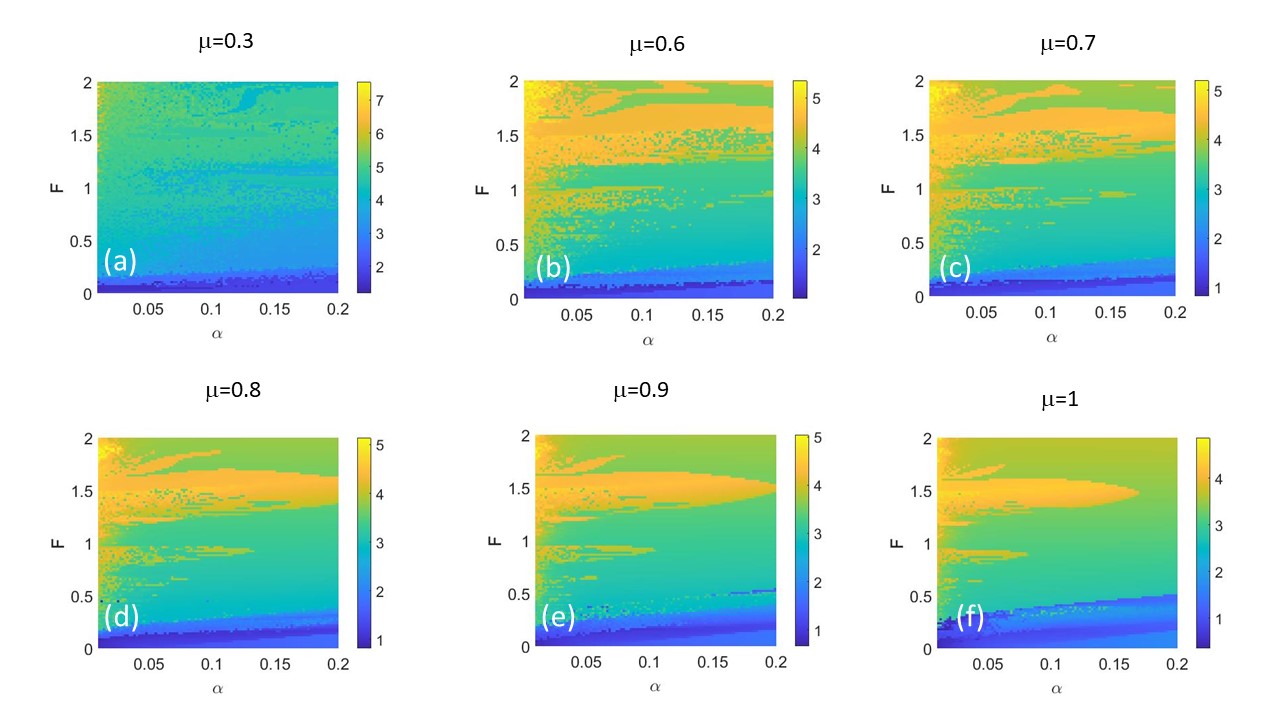}
   \caption{We show the oscillations amplitude in the parameter set $F-\alpha$, for different $\mu$ values. Here,  it appears along the panels the formation of a well located region of  high oscillations amplitude near $F=1.5$  }
\label{fig:4}
\end{figure}

\begin{figure}[htbp]
  \centering
   \includegraphics[width=16.0cm,clip=true]{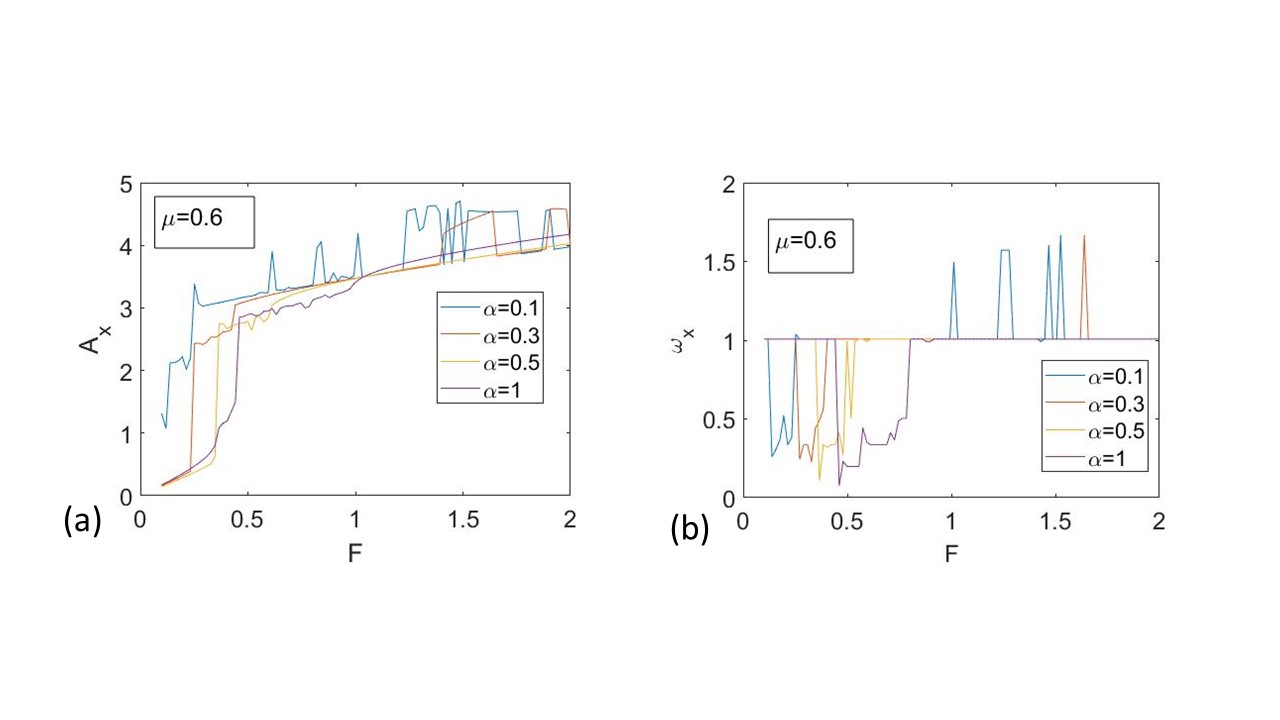}
   \caption{ The figure shows four slices of Fig.~\ref{fig:4}(b) overlapped in panel (a) and the oscillations frequencies calculated with the Fast Fourier transform in panel (b). In the first panel, the curves related to fractional order values, $\alpha=0.1$ and $\alpha=0.3$, that cross the high amplitude oscillations region are recognisable. The other two, after they grow at $F\sim0.5$, become smooth. The frequencies in panel (b) for $F\gtrsim0.9$ tend to the forcing frequency $\omega=1$, except for the ones related with the resonance peak.  }
\label{fig:5}
\end{figure}

\begin{figure}[htbp]
  \centering
   \includegraphics[width=16.0cm,clip=true]{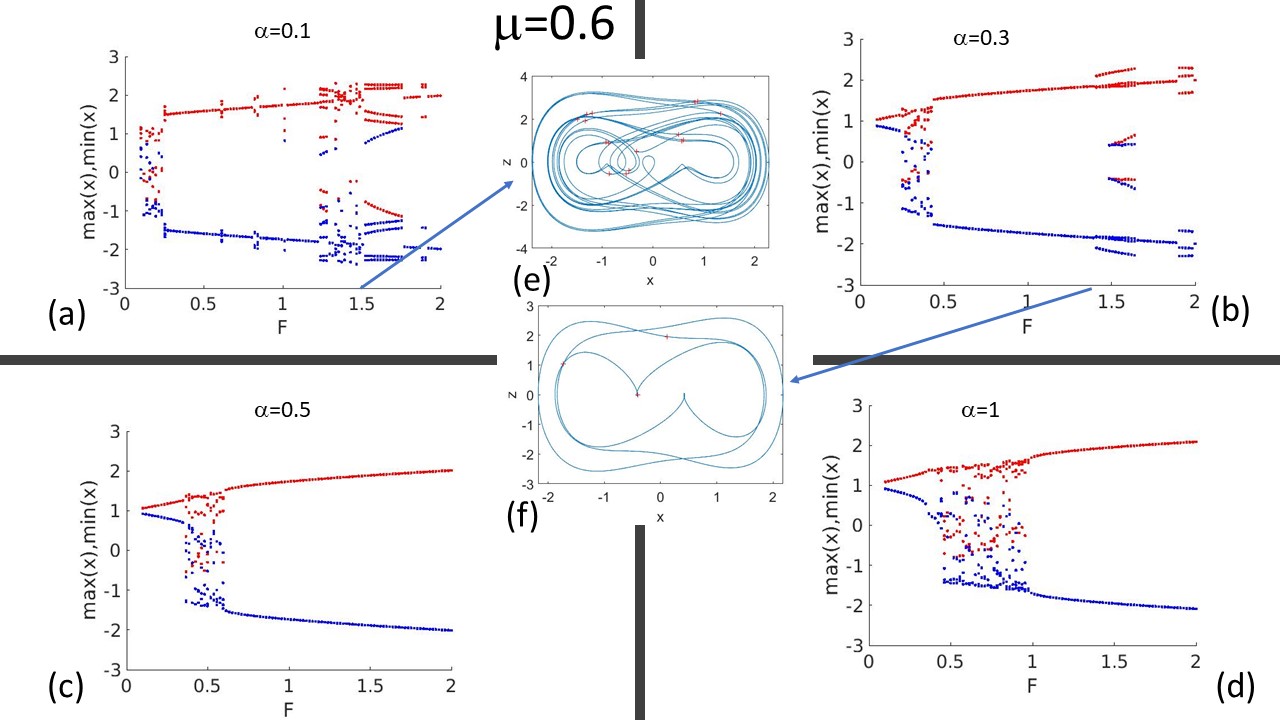}
   \caption{ We show the behavior of the system for fixed values of  $\alpha$ and $\mu$. Comparing panel (a) and (b), we can observe that the system behaviors at $F=1.5$ are different. For $\alpha=0.1$, in panel (e), the oscillations are clearly not periodic, while for $\alpha=0.3$, in panel (f), the oscillations are periodic of period $3$. In the other two panels, the higher oscillations amplitude enlargement is lacking. }
\label{fig:6}
\end{figure}

\section{The forcing amplitude effect on the oscillations amplitude}\label{section_4}

To carry on our analysis, we have decided to study the effect of the forcing amplitude $F$ on the oscillations amplitude since it will provide more easily clear light on the resonance-like phenomenon. The main result of this manuscript is shown in Fig.~\ref{fig:4}, where we have depicted the gradient parameter set $F-\alpha$ of the oscillations amplitude for different values of $\mu$. As expected, the higher the damping term the lower the oscillations amplitude. Interestingly, a high amplitude oscillations region starts to appear when $\mu>0.5$, in particular it becomes more evident  at $\mu\approx 0.6$. This region is centered around $F=1.5$ and spans all or part of the parameter set, depending on the damping parameter. In fact, we can see there that all along the panels of Fig.~\ref{fig:4} that region shrinks along the $\alpha$ axis together with the oscillations amplitude. That means that for a high enough damping parameter value, the high oscillations amplitude  is confined around a certain forcing amplitude value, that is, $F=1.5$ and it is induced by a certain range of fractional order parameter values. These results corroborate what was observed in Fig.~\ref{fig:0}.

Besides, the region shrinks spanning smaller values of the fractional order parameter when the dissipation grows. This phenomenon can be explained as in the previous section. The interaction among high dissipation, certain values of the fractional order parameter and high forcing amplitude triggers a resonant behavior. The interesting fact is that the gain in amplitude is confined to certain $\alpha$ values, so we can define the phenomenon as a resonance induced by the fractional order parameter. The difference with $F=0.1$ is that now the forcing amplitude is large enough to sustain the high oscillations amplitude inside a large region of the parameter set and not just for some particular values of $\alpha$.

For a better understanding and the characterization of the phenomenon, we have deepened its analysis with the other figures below. We want to stress out that all the figures from now on are related with Fig.~\ref{fig:4} and they provide more specific details of the features of the resonance phenomenon. Therefore, we start by  showing in Fig.~\ref{fig:5}(a) the oscillations amplitude versus the forcing amplitude $F$ for $\mu=0.6$. Every curve is associated with a fractional order parameter value as indicated there. It is possible to see that the amplitude curves corresponding with the two values $\alpha=0.1$ and $\alpha=0.3$, for this damping parameter, cross the high oscillations amplitude region of Fig.~\ref{fig:4}(b) and present the corresponding peaks, near $F=1.5$. On the other hand, the curves related with the other two $\alpha$ values are smooth after a first jump, at  $\alpha\approx0.5$.   In Fig.~\ref{fig:5}(b), we show the frequencies of the oscillations, calculated with the Fast Fourier transform. It is possible to see that from $F=0.8$ the calculated frequencies tend naturally to the forcing frequency $\omega=1$, but not near the forcing amplitude for which the resonance peaks are produced for the $\alpha$ values that cross the high amplitude region of Fig.~\ref{fig:4}(b).

In order to have a deeper insight on the dynamics of the system, we plot Fig.~\ref{fig:6} in which we show the maxima-minima diagram for $\mu=0.6$ and for $4$ values of the $\alpha$ parameter, along with two orbits in the case of $F=1.5$ and $\alpha=0.1$ (Fig.~\ref{fig:6}(e)) and $\alpha=0.3$ (Figs.~\ref{fig:6}(f)). These fractional order parameter values are the ones that cross the high oscillations amplitude in Fig.~\ref{fig:4}(b). The other two values, see Figs.~\ref{fig:6}(c) and (d), do not cross that region as indicated by the maxima-minima diagram smoothness.  One consideration can be made, we can see that there is an important difference between the two plotted orbits, as seen in Figs.~\ref{fig:6}(e) and (f). The first one is not periodic, while the second one is periodic of period $3$. It seems that when the fractional order parameter grows, the large amplitude aperiodic trajectories disappear becoming periodic as if the fractional order parameter contributed to introduce order into chaos.

\begin{figure}[htbp]
  \centering
   \includegraphics[width=16.0cm,clip=true]{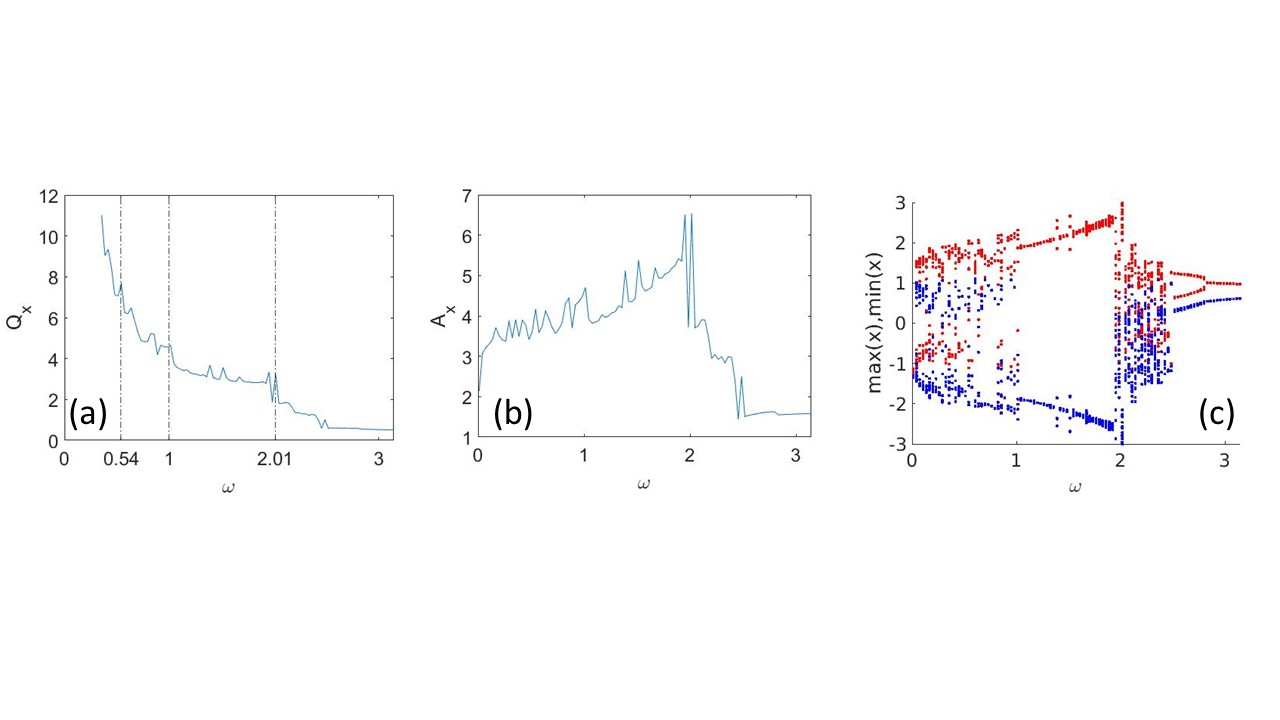}
   \caption{We show the $Q-$factor (a), the oscillations amplitude (b) and the maxima-minima diagram in function of the forcing frequency $\omega$ for the following parameters: $\mu=0.6$, $\alpha=0.1$ and $F=1.5$. In the first panel, three peaks are emphasized with three vertical lines in order to compare them with similar peaks in other forthcoming figures. The peaks in the panels are the result of the equilibrium reached by the high oscillations suppression due to the damping term, the effect of the forcing amplitude and the action of the fractional order parameter.}
\label{fig:7}
\end{figure}

Now, we check the robustness of the resonance induced by the fractional order parameter when the forcing frequency $\omega$ changes. For that purpose,  we plot the $Q-$factor in Fig.~\ref{fig:7}, which shows the oscillations amplitude and the maxima-minima diagram in function of the forcing frequency and for $F=1.5$ and $\mu=0.6$. In Figs.~\ref{fig:7}(a) and (b) it is possible to notice some peaks in the $Q-$factor curve that assure us that they are related with the fractional order parameter induced resonance. This resonance phenomenon needs an interaction among high forcing amplitude, high damping parameter and small values of the fractional order parameter. In Fig.~\ref{fig:7}(a), we have highlighted three peaks at $\alpha=0.54,\alpha=1$ and $\alpha=2.01$ that will be discussed later when we analyze higher $\mu$ values. The peaks are corroborated by the maxima-minima diagram, as seen in Fig.~\ref{fig:7}(c). In this last figure, it is interesting to notice that the higher peaks are related with aperiodic trajectories.

\begin{figure}[htbp]
  \centering
   \includegraphics[width=16.0cm,clip=true]{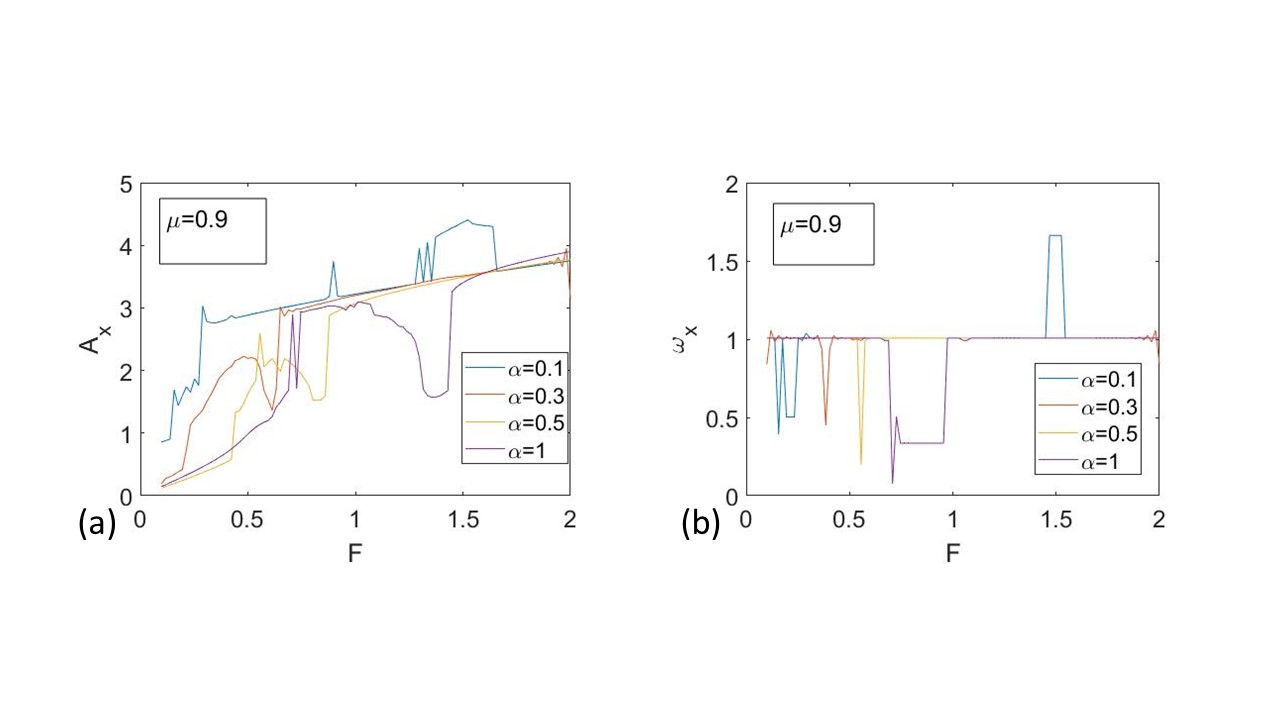}
   \caption{Panel (a) shows four slices of Fig.~\ref{fig:4}(e) overlapped. Panel (b) shows the oscillations frequencies calculated with the fast Fourier transform. In panel (a), it is possible to see that the curve for $\alpha=0.1$ is the only one showing the peak due to the high oscillations amplitude region of Fig.~\ref{fig:6}.  Also here, the oscillations frequencies tend to the forcing frequency $\omega=1$, except for the ones related with the oscillations amplitude peak.   }
\label{fig:8}
\end{figure}

\begin{figure}[htbp]
  \centering
   \includegraphics[width=16.0cm,clip=true]{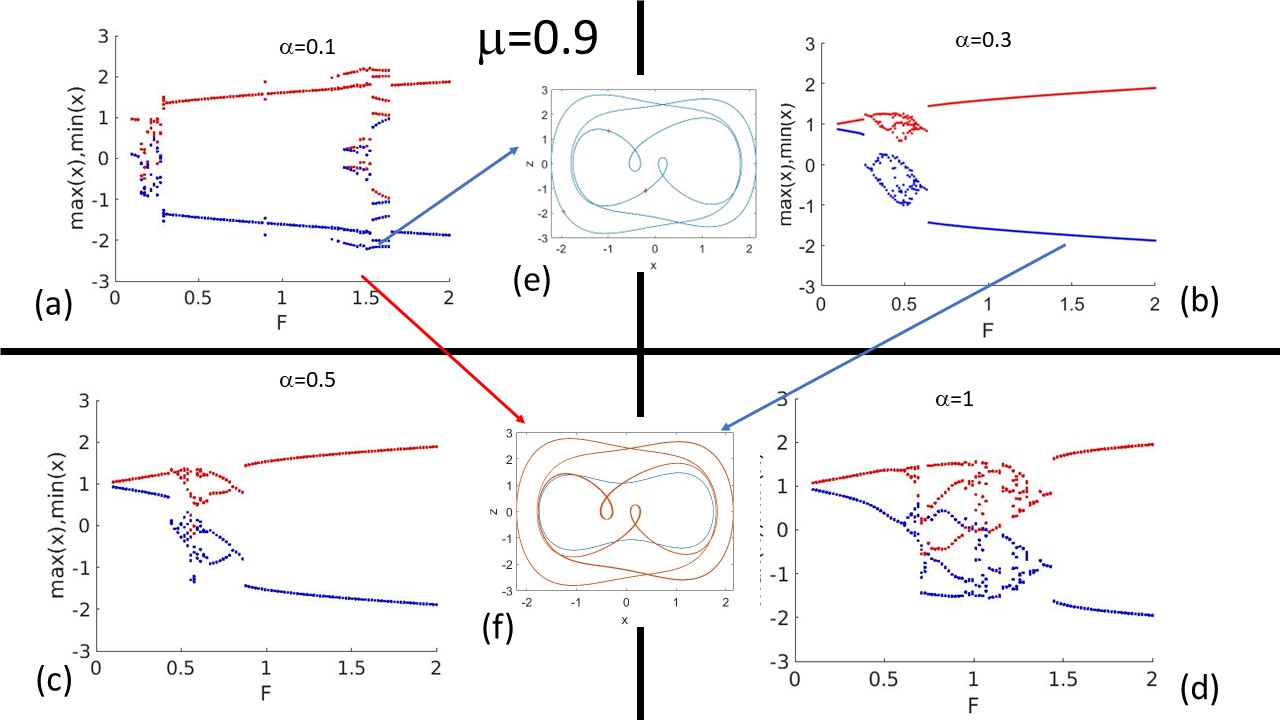}
   \caption{The panels show the behavior of the system for fixed $\alpha$ and $\mu$. Analyzing panel (a), we can see that the system behavior at $F=1.5$ is periodic of period $3$, as shown in panel (e), unlike in Fig.~\ref{fig:6}(e).  In the other three panels, the higher oscillations amplitude  enlargement is lacking. In fact, in panel (f) we compare the orbits for $\alpha=0.1$ and $\alpha=0.3$ and we can appreciate the red trajectory broadening with respect to the blue orbit, due to the resonance effect.}
\label{fig:9}
\end{figure}

\begin{figure}[htbp]
  \centering
   \includegraphics[width=16.0cm,clip=true]{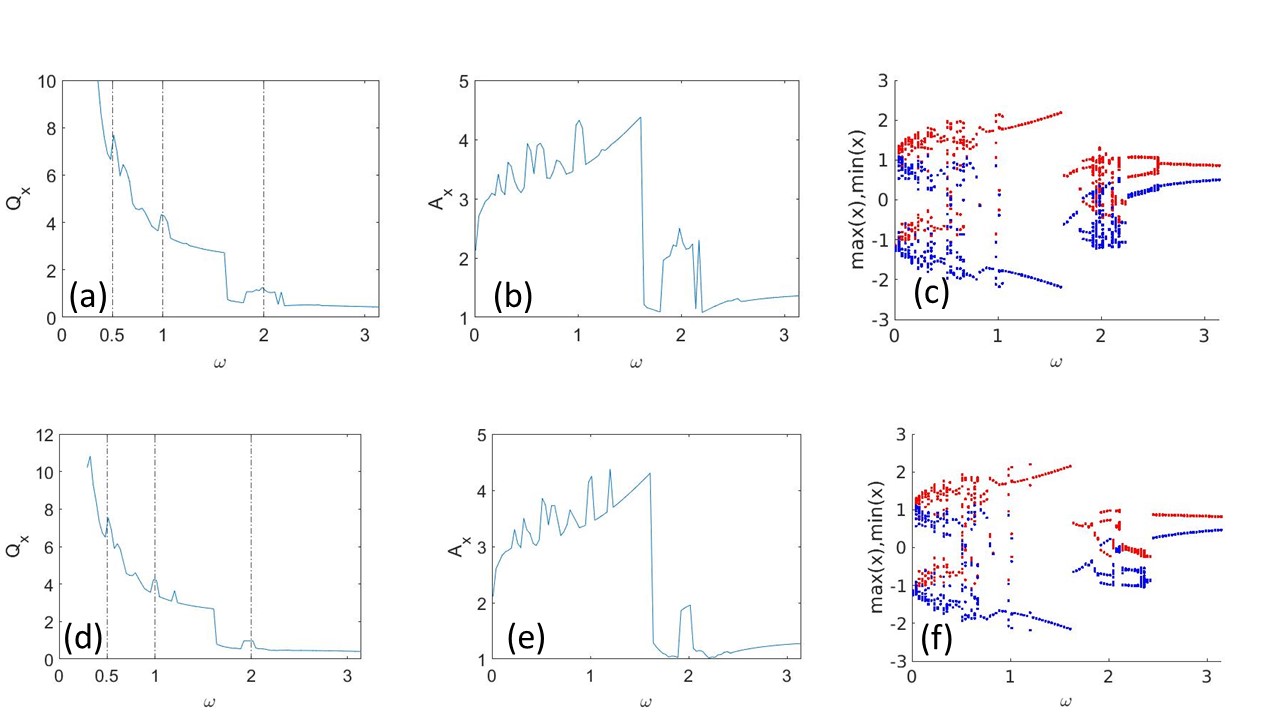}
   \caption{The panels show for $\mu=0.9$, $\alpha=0.1$ and $F=1.5$ (a) the $Q-$factor, (b)  the oscillations amplitude  and (c) the maxima-minima diagram in function of the forcing frequency $\omega$. Panels (d), (e) and (f) show respectively the same plots for $\mu=1$. In panels (a) and (d) the already commented three peaks are emphasized with three vertical lines in order to compare them with similar peaks in Fig.~\ref{fig:7}. Those peaks can be seen more clear, and others, that can be seen in Fig.~\ref{fig:7}, are disappearing as an effect of the dissipation growth. Also they are centered to certain values of the forcing frequency $\omega=0.5, \omega=1$ and $\omega=2$. }
\label{fig:10}
\end{figure}

Now, we analyze the situation in which the damping parameter has a large value and, therefore, we are in the overdamped regime. In Fig.~\ref{fig:8}, we can find a similar plot as Fig.~\ref{fig:5} for the damping parameter value $\mu=0.9$. Here, in Fig.~\ref{fig:8}(a), we can see that only for $\alpha=0.1$ there is a resonance peak, as expected from Fig.~\ref{fig:4}(e). The computed oscillations frequencies, Fig.~\ref{fig:8}(b), behave as in the previous case.  This result is confirmed by Fig.~\ref{fig:9}, where we can see that the resonance amplitude can be found only for $\alpha=0.1$, Fig.~\ref{fig:9}(a), while for the others fractional values the maxima-minima diagram is smooth for large forcing amplitudes, Figs.~\ref{fig:9}(b)-(d). Interestingly,  the high oscillations amplitude  is related with periodic trajectories, as we can see in Fig.~\ref{fig:9}(e), differently from what was happening in Fig.~\ref{fig:6}(e). We also compare the trajectories for $F=1.5$ at $\alpha=0.1$ and at $\alpha=0.3$ in Fig.~\ref{fig:9}(f) and we can appreciate the gain in amplitude, due to the resonance. Then, we plot Fig.~\ref{fig:10}, by varying the forcing frequency $\omega$. Here, we can spot the three peaks already mentioned before, as depicted in Fig.~\ref{fig:7}(a). Again, the interaction between high damping parameter, forcing and fractional order parameter induces the resonance. In this figure, we can see that the $Q-$factor curves, Fig.~\ref{fig:10}(a) and  Fig.~\ref{fig:10}(d), become smoother and that the three peaks at $\omega=0.5,\omega=1$ and $\omega=2$ are more recognizable. Of course, those peaks are related with higher oscillations amplitude, as shown  in Fig.~\ref{fig:10}(b) and Fig.~\ref{fig:10}(e). Then, in Fig.~\ref{fig:10}(c) and Fig.~\ref{fig:10}(f), we can appreciate that the peaks match regions of non periodic dynamics of the system. Contrarily to the $\mu=0.6$ situation, in the $\mu=0.9$ and $\mu=1$ cases, the resonance peaks are centered to specific values of the forcing frequency $\omega$, related among them as $\omega_0=1$, $\omega_{1/2}=\omega_0/2$ and $\omega_2=2\omega_0$. So, we obtained the resonance frequency and its harmonic and subharmonic.  Also, in Fig.~\ref{fig:10}(c) and Fig.~\ref{fig:10}(f), we can see that the dissipation keeps doing its job by bringing more order into chaos.

\begin{figure}[htbp]
  \centering
   \includegraphics[width=16.0cm,clip=true]{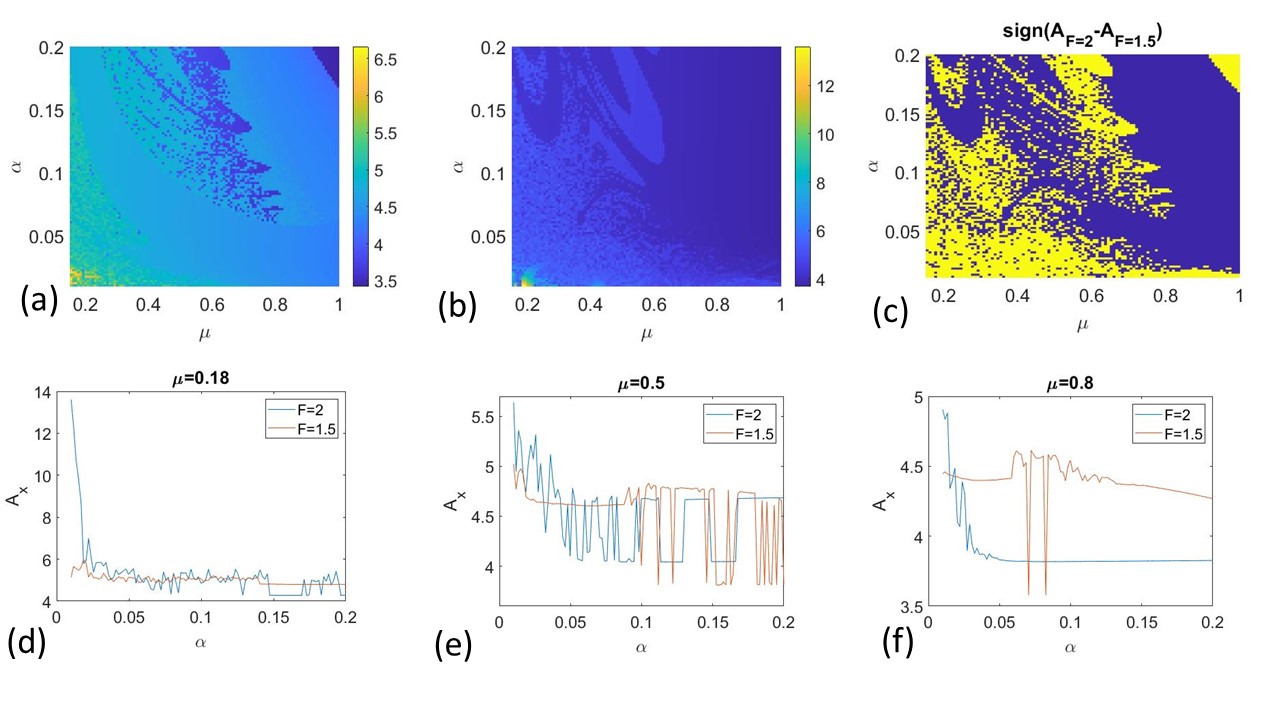}
   \caption{ The figure shows (a) the gradient of the oscillations amplitude in the parameter set $\alpha-\mu$ for $F=1$, (b) the same gradient plot for $F=2$, and (c) the sign of the difference $A_{F=2}-A_{F=1.5}$. In panel (c) the yellow points show the positive sign of the difference, while the blue points represent the negative sign. The panels (d-f) show the oscillations amplitude at different values of the damping parameter $\mu$ and for the forcing amplitudes: $F=2$ in blue and $F=1.5$ in red. We can see that as the damping parameter grows, the oscillations amplitude for $F=1.5$ becomes higher than in the case of $F=2$. }
\label{fig:11}
\end{figure}

Finally, we present the oscillations amplitude gradient for $F=1$ in the parameter set $\mu-\alpha$, Fig.~\ref{fig:11}(a) and for $F=2$ in Fig.~\ref{fig:11}(b). As expected, in the both cases, the high oscillations amplitudes are common in all the parameter set, although the higher ones are concentrated at small $\mu$ and $\alpha$ values, like in Fig.~\ref{fig:2}(b). Then, in the first case, $F=1.5$, the lower amplitude oscillations can be found in the exact opposite corner of the figure and scattered all around its upper center. Otherwise, in the second case, $F=2$, as the damping parameter reaches the value $\mu=0.6$, the oscillations amplitude diminishes abruptly and for the parameter values for which the resonance is triggered the oscillations amplitude becomes smaller than in the $F=1.5$ case. This outcome is corroborated by Fig.~\ref{fig:11}(c), where  the sign of the difference between the oscillations amplitude at $F=2$ and $F=1.5$ is shown. The yellow points indicate a positive difference $A_{F=2}>A_{F=1.5}$, the blue ones a negative difference $A_{F=1.5}>A_{F=2}$. The points in this plot are clearly intermingled, but the blue points generate a more defined region for higher $\mu$ values. This figure shows well that, although the forcing amplitude is bigger in the $F=2$ case, the interaction between the damping parameter and the fractional order parameter can trigger trajectories with a higher oscillations amplitude, giving birth to a resonance for $F=1.5$. Then, in the other panels, we show three slices of the two gradient plots, shown in Fig.~\ref{fig:11}(a) and Fig.~\ref{fig:11}(b). Here, we can see that, for small damping parameter values ($\mu=0.18$) the oscillations amplitude for $F=2$ is generally comparable with the oscillations amplitude for $F=1.5$. When the damping parameter grows, $\mu=0.5$ the oscillations amplitude for $F=1.5$ are a little bit higher than the $F=2$ case, but not for all the $\alpha$ values. Then, when the damping parameter grows further, $\mu=0.8$, the oscillations amplitude for $F=1.5$ is definitively higher. Figures~\ref{fig:11}(d-f) show the tendency of the oscillations amplitude for both forcing amplitudes.  All of this corroborate the results shown in Fig.~\ref{fig:11}(c). Again, at a high damping parameter value the $F=1.5$ case oscillates at higher amplitudes due to the resonance induced by the interaction among the high damping parameter, the fractional order parameter and the forcing amplitude.

\FloatBarrier

\section{Conclusions and discussion}\label{sec:conclusions}
We have studied the fractional Duffing oscillator with a damping fractional term, using the Gr\"unwald-Letnikov integrator. For the parameter values used, we have find that the fractional order parameter can induce a resonance. When the forcing is relatively small, at $F=0.1$, the peaks related with the resonance appear for particular values of the fractional order parameter, values that change in function of the damping parameter.  
For larger values of the forcing amplitude, say $F=1.5$, the higher oscillations amplitude due to the resonance form a region in the parameter set ($F-\alpha$). In both cases the phenomenon is clear, the high oscillations amplitude, that the oscillator exhibits at low damping parameter values, are attenuated at higher dissipation but the interaction between a particular forcing amplitude and certain low fractional order parameter values maintain them, generating a resonance.  We think that this phenomenon can be relevant to better understand the relation between the damping parameter and the fractional order parameter. Besides, it shows the richness in the dynamical behavior due to the presence of a fractional term in the nonlinear oscillator. This last point can easily be observed in the overdamped case, where the dynamics of the system becomes still complex. Finally, we expect that this study can be useful to  predict the birth of complex or resonant behaviors in viscoelastic and porous materials or diffusion processes, that can be only modeled by using fractional derivatives.

\section{Acknowledgments}

This work has been supported by the Spanish State Research Agency (AEI) and the European Regional Development Fund (ERDF, EU) under Project No.~PID2019-105554GB-I00 (MCIN/AEI/10.13039/501100011033).


\begin{thebibliography}{99}
\bibitem{Boroviec} Borowiec M, Litak G, Syta A.  Vibration of the Duffing oscillator: effect of fractional damping. Shock Vib. 2007; 14: 29-36.
\bibitem{Dafermos} Dafermos C. Dissipation in materials with memory. In:  Lodge AS,  Renardy M,  Nohel JA, editors. Viscoelasticity and Rheology. Proceedings of a Symposium Conducted by the Mathematics Research Center,  Wisconsin–Madison: Academic Press; 1985, p. 221-234.
\bibitem{Lu} Lu Z, Wang Z, Zhou Y, Lu X.  Nonlinear dissipative devices in structural vibration control: A review. J. Sound Vib. 2018; 423: 18-49.
\bibitem{Chellaboina} Chellaboina V, Haddad WM.  Exponentially dissipative nonlinear dynamical systems: a nonlinear extension of strict positive realness. In: Jeltsema D,  Scherpen J, editors. Proceedings of the 2000 American Control Conference, Chicago: IEEE Xplore; 2000, p. 3123-3127.
\bibitem{De} De S, Kunal K,  Aluru NR.  Nonlinear intrinsic dissipation in single layer MoS 2 resonators. RSC Adv. 2017; 7: 6403-6410.
\bibitem{Elliott} Elliott SJ, Tehrani MG, Langley RS.  Nonlinear damping and quasi-linear modelling. Philos. Trans. R. Soc. A 2015; 373: 20140402.
\bibitem{Horr}Horr AM, Schmidt LC.  A fractional-spectral method for vibration of damped space structures. Eng. Struct. 1996; 18: 947-956.
\bibitem{Ding} Ding Y, Liu X, Chen P, Luo X, Luo Y. Fractional-Order Impedance Control for Robot Manipulator. Fractal fract. 2022; 6: 684.
\bibitem{Li}Li X, Liu Z, Li J, Tisdell C.  Existence and controllability for nonlinear fractional control systems with damping in Hilbert spaces. Acta Math. Sci. 2019; 39; 229-242.
\bibitem{Strogatz}Strogatz SH.  Nonlinear dynamics and chaos with student solutions manual: With applications to physics, biology, chemistry, and engineering. Boca Raton: CRC press; 2018.
\bibitem{Sanjuan} Rajasekar S, Sanju\'an MAF.  Nonlinear resonances. Switzerland: Springer International Publishing; 2016.
\bibitem{Coccolo_fr2}
Coccolo M, Seoane JM, Lenci S, Sanju\'an MAF. Fractional damping effects on the transient dynamics of the Duffing oscillator. Commun. Nonlinear Sci. Numer. Simul. 2023; 117: 106959.
\bibitem{Coccolo_fr} Ortiz A, Yang J, Coccolo M, Seoane JM, Sanju\'an MAF. Fractional damping enhances chaos in the nonlinear Helmholtz oscillator. Nonlinear Dyn. 2020; 102: 2323-2337.
\bibitem{Syta} Syta A, Litak G, Lenci S, Scheffler M.  Chaotic vibrations of the duffing system with fractional damping.  Chaos 2014; 24: 013107.


\end{thebibliography}
\end{document}